\documentclass[a4paper,UKenglish,cleveref, autoref, thm-restate]{lipics-v2021}

\usepackage{graphicx,listings,tikz}
\usepackage{algorithm,algorithmicx,algpseudocode}
\MakeRobust{\Call}

\usepackage{parcolumns}
\usepackage{listings}
\newcommand{\mycommentstyle}{\normalfont\itshape}

\usepackage{threeparttable,tabularx,multirow,multicol}
\usepackage{wrapfig}
\usepackage{float}
\newcommand{\Cerny}{\v{C}ern{\'y} }
\usepackage{mathrsfs}
\DeclareSymbolFont{rsfscript}{OMS}{rsfs}{m}{n}
\DeclareSymbolFontAlphabet{\mathrsfs}{rsfscript}
\usepackage{anyfontsize}
\renewcommand{\O}{\mathcal{O}}

\algnewcommand\algorithmicswitch{\textbf{switch}}
\algnewcommand\algorithmiccase{\textbf{case}}
\algnewcommand\algorithmicassert{\texttt{assert}}
\algnewcommand\Assert[1]{\State \algorithmicassert(#1)}%
\algdef{SE}[SWITCH]{Switch}{EndSwitch}[1]{\algorithmicswitch\ #1\ \algorithmicdo}{\algorithmicend\ \algorithmicswitch}%
\algdef{SE}[CASE]{Case}{EndCase}[1]{\algorithmiccase\ #1}{\algorithmicend\ \algorithmiccase}%
\algtext*{EndSwitch}%
\algtext*{EndCase}%

\title{An Improved Algorithm for Finding the Shortest Synchronizing Words}

\author{Marek Szyku{\l}a}{University of Wroc{\l}aw, Faculty of Mathematics and Computer Science, Poland}{msz@cs.uni.wroc.pl}{https://orcid.org/0000-0001-5349-468X}{}
\author{Adam Zyzik}{University of Wroc{\l}aw, Faculty of Mathematics and Computer Science, Poland}{adamkzyzik@gmail.com}{}{}

\authorrunning{M. Szyku{\l}a, A. Zyzik} 

\Copyright{Marek Szyku{\l}a and Adam Zyzik} 



\category{} 

\relatedversion{} 

\funding{This work was supported in part by the National Science Centre, Poland under project number 2021/41/B/ST6/03691.
The experiments were run on a computational grid of the Institute of Computer Science, University of Wroc{\l}aw, funded by National Science Centre, Poland, under project number 2019/35/B/ST6/04379.
}

\nolinenumbers 

\EventEditors{John Q. Open and Joan R. Access}
\EventNoEds{2}
\EventLongTitle{42nd Conference on Very Important Topics (CVIT 2016)}
\EventShortTitle{CVIT 2016}
\EventAcronym{CVIT}
\EventYear{2016}
\EventDate{December 24--27, 2016}
\EventLocation{Little Whinging, United Kingdom}
\EventLogo{}
\SeriesVolume{42}
\ArticleNo{23}
\begin{document}
\maketitle
\begin{abstract}
A synchronizing word of a deterministic finite complete automaton is a word whose action maps every state to a single one.
Finding a shortest or a short synchronizing word is a central computational problem in the theory of synchronizing automata and is applied in other areas such as model-based testing and the theory of codes.
Because the problem of finding a shortest synchronizing word is computationally hard, among \emph{exact} algorithms only exponential ones are known.
We redesign the previously fastest known exact algorithm based on the bidirectional breadth-first search and improve it with respect to time and space in a practical sense. We develop new algorithmic enhancements and adapt the algorithm to multithreaded and GPU computing.
Our experiments show that the new algorithm is multiple times faster than the previously fastest one and its advantage quickly grows with the hardness of the problem instance.
Given a modest time limit, we compute the lengths of the shortest synchronizing words for random binary automata up to 570 states, significantly beating the previous record.
We refine the experimental estimation of the average reset threshold of these automata.
Finally, we develop a general computational package devoted to the problem, where an efficient and practical implementation of our algorithm is included, together with several well-known heuristics.
\end{abstract}
\section{Introduction}

A \emph{deterministic finite complete semi-automaton} (called simply an \emph{automaton}) is a $3$-tuple $(Q,\Sigma,\delta)$, where $Q$ is a finite set of \emph{states}, $\Sigma$ is an \emph{input alphabet}, and $\delta\colon Q \times \Sigma \to Q$ is a completely defined \emph{transition function}.
The transition function is naturally extended to a function $Q \times \Sigma^* \to Q$.
Throughout the paper, by $n$ we denote the number of states in $Q$ and by $k$ we denote the size of the input alphabet $\Sigma$.
A word is \emph{reset} (or \emph{synchronizing}) if $|\delta(Q,w)|=1$; in other words, for every two states $p,q \in Q$ we have $\delta(q,w) = \delta(p,w)$.
An automaton that admits a reset word is called \emph{synchronizing}.

The classical synchronization problem is, for a given synchronizing automaton, to find a reset word.
Preferably, this word should be as short as possible.
Therefore, the main property of a synchronizing automaton is its \emph{reset threshold}, which is the length of the shortest reset words.
We denote the reset threshold by $r$.
Synchronizing automata and the synchronization problem are known for both their theoretical properties and practical applications.

\subsection{Theoretical Developments}

On the theoretical side, there is a famous long-standing open problem from 1969 called the \Cerny conjecture; see an old \cite{Volkov2008Survey} and a recent survey \cite{KariVolkov2021Survey}.
The conjecture claims that the reset threshold is at most $(n-1)^2$.
If true, the bound would be tight, as the \Cerny automata meet the bound for each $n$ \cite{Cerny1964}.

Until 2017, the best known upper bound on the reset threshold was $(n^3-n)/6-1 \sim 0.1666\ldots n^3 + \O(n^2)$ ($n \ge 4)$ \cite{Pin1983OnTwoCombinatorialProblems} by the well-known Frankl-Pin's bound.
The current best known upper bound is $\sim 0.1654 n^3 + o(n^3)$ by Shitov \cite{Shitov2019}, which was obtained by refining the previous improvement $\sim 0.1664 n^3 + \O(n^2)$ by Szyku{\l}a \cite{Szykula2018ImprovingTheUpperBound}.
Apart from that, better bounds were obtained for many special subclasses of automata.
Synchronizing automata are also applied in other theoretical areas, e.g., matrix theory \cite{GGJ2018PrimitiveSets}, theory of codes \cite{BPR2010CodesAndAutomata}, Markov processes \cite{TC2011ExactSynchronization}.
Several new results around the topic appear every year.
Recently, a special journal issue was dedicated to the problem \cite{JALCCerny2019} for the occasion of the 50th anniversary of the problem.

Reset thresholds were also studied for the average case.
Berlinkov showed that a random binary automaton is synchronizing with high probability \cite{Berlinkov2016OnTheProbabilityToBeSynchronizable}.
Moreover, Nicaud showed that such an automaton with high probability has a reset threshold in $\O(n \log^3 n)$ \cite{Nicaud2014FastSynchronizationOfRandomAutomata}.
Based on that, the upper bound $\O(n^{3/2+o(1)})$ on the expected reset threshold of a random binary automaton was obtained \cite{BerlinkovSzykulaS2016AlgebraicSynchronizationCriterion}.
These studies were accompanied by experiments, and the best estimation obtained so far was $2.5 \sqrt{n-5}$ \cite{KKS2013FastAlgorithm}.

There were also performed massive experiments directly aimed at verifying the \Cerny conjecture and other theoretical properties for automata with small numbers of states \cite{BDZ2019LowerBoundsInPartialAutomata,KKS2016ExperimentsWithSynchronizingAutomata,Trahtman2006Trends}.
For all such studies, finding (the length of) a shortest reset word is a crucial problem.

\subsection{Synchronization in Applications}

Apart from the theory, the synchronization problem finds applications in practical areas, e.g., testing of reactive systems \cite{Pomeranz1994,Sandberg2005Survey}, networks \cite{Kari2002}, robotics \cite{AV2003SynchronizingMonotonicAutomata}, and codes \cite{Jurgensen2008}.

Automata are frequently used to model the behavior of systems, devices, circuits, etc.
The idea of synchronization is natural: we aim to restore control over a device whose current state is not known or we do not want our actions to be dependent on it.
For instance, for digital circuits, where we need to test the conformance of the system according to its model, each test is an input word and before we run the next one, we need to restart the device.
In another setting, we are an observer who knows the structure of the automaton but does not see its current state and wants to eventually learn it by observing the input; once a reset word appears, the state is revealed unambiguously.
See a survey \cite{Sandberg2005Survey} explaining synchronizing sequences and their generalization to automata with output: \emph{homing sequences}, which\ allow determining the (hidden) current state of the automaton by additionally observing the generated automaton's output.

Another particular application comes from the theory of codes, where finite automata act as \emph{decoders} of a compressed input.
Synchronizing words can make a code resistant to errors, since if an error occurs, after reading such a word, decoding is restored to the correct path.
See a book for the role of synchronization in the theory of codes \cite{BPR2010CodesAndAutomata} and recent works \cite{BFRS21SynchronizingStronglyConnectedPartialDFAs} explaining synchronization applied to prefix codes.

\subsection{Algorithms Finding Reset Words}

Determining the reset threshold is computationally hard.
The decision problem, whether the reset threshold is smaller than a given integer, is NP-complete \cite{Eppstein1990}, and it remains hard even for very restrictive classes such as binary Eulerian automata \cite{V2017ComplexityEulerian}.
The functional problems of computing the reset threshold and a shortest reset word are respectively $\mathrm{FP}^{\mathrm{NP}[\log]}$-complete and $\mathrm{FP}^{\mathrm{NP}}$-complete \cite{OlschewskiUmmels2010}.
Moreover, approximating the reset threshold is hard even for approximation factors in $\O(n^{1-\varepsilon})$, for every $\varepsilon > 0$ \cite{GawrychowskiStraszak2015StrongInapproximability}.
This inapproximatibility also holds for subclasses related to prefix codes \cite{RS2018FindingSynchronizingWordsForPrefixCodes}.
On the other hand, there exists a simple general $\O(n)$-approximating polynomial algorithm \cite{GerbushHeeringa2011}.
It is open whether there exists a polynomial algorithm approximating within some sublinear factor, e.g., in $\O(n/\log n)$.

From the perspective of fixed-parameter tractability, the main parameter determining the hardness is the reset threshold itself (and the alphabet size, if not fixed).
It plays a similar role as the number of variables in the SAT problem, yet, in contrast, it is not given but is the result to be computed.
It is trivial to compute the reset threshold in time $\O(n \cdot k^r)$ simply by checking all words of length $1, 2, \ldots, r$ (we need $\O(n)$ time for computing the image of a subset of $Q$ under the action of one letter).
This essentially cannot be much better, as assuming the Strong Exponential Time Hypothesis (SETH), the problem cannot be solved in time $\O^*((k-\varepsilon)^r)$, for every $\varepsilon > 0$ \cite[Theorem~8]{FHV2015AMultiParameterAnalysisOfHardProblemsOnDFAs}, where $\O^*$ suppresses all polynomial factors in the size of the input.

Therefore, exponential exact algorithms that hopefully find a shortest reset word faster in typical or average cases are used.
Alternatively, there are many polynomial heuristics proposed that find a relatively short reset word in practice.

\subsubsection{Exact Algorithms}

In general, exact algorithms can be used for automata that are not too large.
They also play an important role in testing heuristics, providing the baseline for comparison (e.g., \cite{RS2015ForwardAndBackward}).

The naive algorithm of checking all words is practically slow, as it does not involve any optimization and works always in time $\O^*(k^r)$.
The standard algorithm, e.g., \cite{KudlacikRomanWagner2012,Sandberg2005Survey,Trahtman2006Trends}, for computing a shortest reset word is finding a path of state subsets in the power automaton (that is, the automaton whose set of states is $2^Q$) from $Q$ to a singleton.
This works in at most $\O(kn\cdot 2^n)$ time but is practically faster as we usually traverse through fewer sets.
Note that $n$ may be much smaller than $r$  (we know examples where $r$ can be quadratic in $n$, e.g., \cite{AGV2013SlowlySynchronizing}), but in the average case it is the opposite.
The main drawback of this algorithm is its requirement of $\O(2^n)$ space, which is acceptable only up to small $n \sim 30$.

Alternative approaches include utilizing SAT solvers \cite{SkvortsovTipikin2011} by suitable reductions of the problem and binary search over possible values of $r$.
SAT solvers were also recently tried for partial deterministic finite automata and \emph{careful synchronization} \cite{Shabana2019}.
In the reported results, such solutions reach random binary automata with about $100$ states.

The fastest algorithm so far is based on a bidirectional search of the power automaton, equipped with several enhancements \cite{KKS2013FastAlgorithm,KKSS2015ComputingTheShortestResetWords}.
This algorithm was able to deal with binary random automata up to $350$ states.
Despite several later attempts, no faster solutions were developed and the algorithm was not improved until now.

\subsubsection{Heuristic Algorithms}

The most classic polynomial algorithm is Eppstein's one \cite{Eppstein1990}.
It works in $\O(n^3 + k n^2)$ time and finds a reset word of length at most $(n^3-n)/3$ (due to the Frankl-Pin's bound).
Several heuristic improvements were proposed, e.g., Cycle, SynchroP, and SynchroPL algorithms \cite{KudlacikRomanWagner2012,Trahtman2006Trends}, which do not improve guarantees but behave better in experimental settings, even at the cost of increased worst-case time complexity.
Recent works also involve attempts to speed-up heuristics by adapting to parallel and GPU computation \cite{SAAKY2021BoostingExpensiveSynchronizingHeuristics,TKY2021SynchronizingBillionScaleAutomata}.

A remarkable heuristic is the \emph{beam} algorithm based on inverse breadth-first search (\cite[\emph{CutOff-IBFS}]{RS2015ForwardAndBackward}), i.e., starting from a singleton and ending with $Q$, which significantly beats other algorithms based on the forward search.
Yet, curiously, it does not provide any worst-case guarantees, as it theoretically may not find any reset word at all.

Alternative approaches involve artificial intelligence methods, e.g., hierarchical classifier \cite{PRJ2012ApplicationOfHierarchicalClassifier}, genetic algorithms \cite{KR2017NewEvolutionaryAlgorithmForSynchronization}, and machine learning approaches \cite{PRSZ2018MachineLearningSynchronization}.

\subsection{Contribution}

We reinvestigate the so-far best exact algorithm \cite{KKS2013FastAlgorithm,KKSS2015ComputingTheShortestResetWords} and significantly improve it.
We develop a series of algorithmic enhancements involving better data structures, decision mechanisms, and reduction procedures.
Altogether, we obtain a significant speed-up and decrease the memory requirements.
Additionally, the remodeled algorithm is adapted for effective usage of multithreading and GPU computing, which was not possible in the original.

On the implementation side, we develop an open computational package containing the new exact algorithm as well as several known polynomial heuristics.
We apply a series of technical optimizations and fine-tune the algorithm to maximize efficiency.
The package supports configurable just-in-time compiled computation plans and can be extended with new algorithms.

In the experimental section, we test the efficiency of the algorithm and compute the reset thresholds of binary random automata up to 570 states.
We refine the previous estimation formula for the expected reset threshold of these automata.

The computational package is available at~\cite{SZ2022Synchrowords} (the version related to this paper is 1.1.0).

\section{The New Exact Algorithm}

Our algorithm is based on the former best exact algorithm
\cite{KKS2013FastAlgorithm,KKSS2015ComputingTheShortestResetWords}.
While the new version differs in the choice of data structures and subprocedures, at a high level it is similar and uses two main phases -- bidirectional breadth-first search and then inverse depth-first search.

The input to the algorithm is an automaton $\mathrsfs{A} = (Q,\Sigma,\delta)$ with $n$ states and $k$ input letters.
The goal is to find its reset threshold $r$.
In the first step, we check if $\mathrsfs{A}$ is synchronizing by the well-known procedure \cite{Volkov2008Survey}, which checks for every two states $p,q \in Q$ whether they can be mapped to one state; this is doable in $\O(kn^2)$.
Then we get upper bounds on the reset threshold by using polynomial-time heuristics.
For this, we use the Eppstein algorithm \cite{Eppstein1990} at first, and then the enhanced \emph{beam} algorithm \cite[CutOff-IBFS]{RS2015ForwardAndBackward}.
The found upper bound helps the main procedure make better decisions.

Given a~subset $S \subseteq Q$ and a word $w \in \Sigma^*$, the \emph{image} of $S$ under the action of $w$ is $\delta(S,w) = \{\delta(q,w) \mid q \in S\}$.
The \emph{preimage} of $S$ under the action of $w$ is $\delta^{-1}(S,w) = \{q \in Q \mid \delta(q,w) \in S\}$.

The key idea is to simultaneously run a breadth-first search (BFS) starting from the set $Q$ and computing images, together with an inverse breadth-first search (IBFS) starting from all of the singletons and computing preimages.
While both algorithms on their own require computation of at most $k^{r}$ or at most $n k^{r}$ sets respectively, combining them lets us compute no more than $n k^{r / 2}$ sets, provided that we can somehow test if the searches have met.
To do this, we need to check if there exists a pair $X$, $Y$ of sets, belonging respectively to the BFS and IBFS lists, such that $X \subseteq Y$.
Indeed, then we know that there are words $x, y \in \Sigma^*$ such that $X = \delta(Q, x)$ and $Y = \delta^{-1}(\{q\}, y)$ for some $q \in Q$.
Because $X \subseteq Y$, we get $\delta(Q, xy) = \{q\}$, which means that $xy$ is a reset word.
Due to the \textit{Orthogonal Vectors Conjecture} \cite{KW2017OVC}, there is probably no subquadratic solution to this subset problem\footnote{The \textit{Orthogonal Vectors problem} gives two sets $A$, $B$ of Boolean vectors of the same length and asks if there exists a pair $(u \in A, v \in B)$ such that $u$ and $v$ are orthogonal, i.e., $u \cdot v = \mathbf{0}$. We can reduce our problem to \textit{OV} by transforming the sets in $Y$ to their complements and then representing all the sets as their characteristic vectors.}.
Such a solution would also contradict the mentioned fact that we cannot find the reset threshold in $\O^*((k-\varepsilon)^r)$ time assuming SETH.
Nevertheless, we employ procedures that work well in our practical case.
They are also used to reduce the number of sets in the lists during the searches, which effectively lowers the branching factor.
Finally, we do not actually run the two searches until they meet.
Instead, we switch to the second phase with inverse DFS (which takes the steps only on the IBFS side computing preimages), when either the memory runs out or we calculate that it should be faster based on the upper bound from the heuristics and the collected statistics.

These high-level ideas are derived from the previous algorithm.
However, we design different, more efficient procedures and optimizations for these steps so that it is possible to solve the problem significantly faster and for larger automata.
First, we modify data structures and redesign how a single iteration of BFS / IBFS works.
Apart from making the bidirectional-search phase faster, it allows completing more iterations before switching to the DFS phase due to the lower memory consumption, which is crucial in the case of large automata.
The decision-making part of the algorithm is also extended.
We use five types of steps, and the decision on which step to take is based on statistics from the current algorithm's run and forward prediction.
In the DFS phase, we enhance the radix trie data structure in terms of both efficiency and memory overhead.
We also apply some forms of list reductions, which decrease the branching factor.
Finally, every part of the new algorithm can be parallelized in one way or the other, which was not possible before; the general difficulty of parallelization comes from large shared data structures and a lot of branching.

In the next sections, we describe the new algorithmic techniques.
For the sake of brevity, we consider a version that only calculates the reset threshold.
The algorithm can be trivially modified to also return the reset word by storing pointers to predecessors along with the sets, although then either time or memory footprint is slightly increased.

\subsection{Bidirectional Breadth-First Search}

The first phase of the algorithm consists of running the two breadth-first searches.
The BFS starts with a list $L_{\mathrm{BFS}}$ containing just the set $Q$.
When the search starts a new iteration, $L_{\mathrm{BFS}}$ is replaced with $\{\delta(S, a) \mid S \in L_{\mathrm{BFS}}, a \in \Sigma\}$.
Conversely, $L_{\mathrm{IBFS}}$ is initialized with all the singletons and the list is replaced with $\{\delta^{-1}(S, w) \mid S \in L_{\mathrm{IBFS}},\, w \in \Sigma\}$.

We say that the two searches \emph{meet} if there exist $X \in L_{\mathrm{BFS}}$ and $Y \in L_{\mathrm{IBFS}}$ for which $X \subseteq Y$ holds.
The meet condition implies that the lists can be reduced by removing the elements which are not
minimal (and respectively maximal for IBFS) with respect to inclusion.
We can reduce the lists further by ensuring that no new set is a superset (subset for IBFS) of a set belonging to some list from any previous iteration.
To make this possible, we keep track of all the visited sets in two additional \textit{history} lists $H_{\mathrm{BFS}}$, $H_{\mathrm{IBFS}}$.
This reduction, although usually helpful during most of the iterations, at the end may turn out to be unprofitable, in which case the algorithm will drop the history list(s).

The original idea for the subprocedure to check the meet condition was to keep the lists as dynamic \textit{radix tries}, supporting insertion and subset (or superset) checking operations.
Now, instead, we take a somewhat simpler approach and operate directly on the lists, stored as random access containers (such as vectors in C++).
We call this subprocedure $\mathit{MarkSupersets}(A, B)$ (and a similar one -- $\mathit{MarkProperSupersets}(A, B)$, which additionally restricts the marked supersets to be non-equal to their subsets).

We split the reductions into three subprocedures: removing duplicates, self-reduction, and then reduction by history.
In contrast to performing only one and the most expensive reduction by history (which could also include the first two reductions), after each subprocedure the list size gets smaller, which makes the next one run faster.

Alg.~\ref{alg:bibfs} shows the pseudocode of the bidirectional-search phase.

\begin{algorithm}[htb!]
\caption{Bidirectional breadth-first search.}\label{alg:bibfs}
\begin{algorithmic}[1]
\Require A synchronizing automaton $\mathrsfs{A}=(Q,\Sigma,\delta)$ with $n = |Q|$ states and $k = |\Sigma|$ input letters. An upper bound $R$ on the reset threshold.
\Ensure Reset threshold $r$.
\State $L_{\mathrm{BFS}}, H_{\mathrm{BFS}} \gets \{Q\}$
\State $L_{\mathrm{IBFS}}, H_{\mathrm{IBFS}} \gets \{\{q\} \mid q \in Q\}$
\For{$r$ \textbf{from} $1$ \textbf{to} $R-1$}
	\Switch{\Call{$\mathit{CalculateBestStep}$}{$ $}}
		\Case{$\mathrm{BFS}$}
		    \If{$H_{\mathrm{BFS}}$ has grown significantly since its last reduction}
			    \State Delete subsets from $H_{\mathrm{BFS}}$ that are larger than the largest ones from $L_{\mathrm{BFS}}$
			    \State Delete non-minimal subsets from $H_{\mathrm{BFS}}$ ($\mathit{MarkProperSupersets}$)
			\EndIf
			\State $L_{\mathrm{BFS}} \gets$ \Call{$\mathit{CalculateImages}$}{$L_{\mathrm{BFS}}$}
			\State Delete duplicates from $L_{\mathrm{BFS}}$ (lex.\ sort)
			\State Delete non-minimal subsets from $L_{\mathrm{BFS}}$ ($\mathit{MarkProperSupersets}$)
			\State Delete supersets of $H_{\mathrm{BFS}}$ from $L_{\mathrm{BFS}}$ ($\mathit{MarkSupersets}$)
  			\State $H_\mathrm{BFS} \gets H_\mathrm{BFS} \cup L_\mathrm{BFS}$
		\EndCase
		\Case{$\mathrm{BFS}^\mathrm{NH}$ (without history)}
			\State $L_{\mathrm{BFS}} \gets$ \Call{$\mathit{CalculateImages}$}{$L_{\mathrm{BFS}}$}
			\State Delete duplicates from $L_{\mathrm{BFS}}$ (lex.\ sort)
			\State Delete non-minimal subsets from $L_{\mathrm{BFS}}$ ($\mathit{MarkProperSupersets}$)
		\EndCase
		\Case{$\mathrm{IBFS}$}
			\State ... \Comment{Analogous to BFS}
		\EndCase
		\Case{$\mathrm{IBFS}^\mathrm{NH}$ (without history)}
			\State ... \Comment{Analogous to BFS (without history)}
		\EndCase
		\Case{DFS}
			\State \Call{$\mathit{DFS}$}{\Call{$\mathit{BuildStaticTrie}$}{$L_{\mathrm{BFS}}$}, $L_{\mathrm{IBFS}}$, $r$, $R$}
			\State \Return $R$ \Comment{$\mathit{DFS}$ sets $R \gets$ the reset threshold}
		\EndCase
  	\EndSwitch

	\If{\Call{$\mathit{MarkSupersets}$}{$L_{\mathrm{BFS}}, L_{\mathrm{IBFS}}$} has found at least one superset}
		\State \Return $r$
	\EndIf
\EndFor
\State \Return $R$
\end{algorithmic}
\end{algorithm}

\subsubsection{Subset and Superset Checking}

There exist several algorithms solving the extremal sets problem in practical settings, e.g., \cite{BP2011FastAlgorithmsForFindingExtremalSets}.
They take a list of sets and mark all those that are not a subset of any other set.

We use a similar method to those utilizing a lexicographic sort, but we operate on two lists and are allowed to change the order of the second list during computation.
$\mathit{MarkSupersets}$ (Alg.~\ref{alg:marksuper}) takes lists $A$ and $B$ and swaps sets in $B$ so that those sets that are supersets of some sets from $A$ appear at the end.
The sets are treated like binary strings, i.e., their characteristic vectors, of length $n$.
The procedure recursively splits the sets in $A$ into those containing the $d$-th state and those not containing it, where $d$ is the recursion depth.
In this sense, it works by implicitly building a radix trie on $A$.
We require that $A$ is sorted lexicographically and its elements are unique, which can be guaranteed relatively cheaply before calling the procedure, as sorting is much faster than subset checking.
The order gives us the property that the sets containing the $d$-th state and those not containing it are stored in continuous segments, so we can effectively split $A$.
This lets us simulate in-place trie traversal with a recursive procedure that takes intervals of the lists as inputs.
When the intervals are small (determined by the constant parameter $\mathit{MIN}$, Alg.~\ref{alg:marksuper} line~2), we can use a brute-force check instead of recursing further, which makes the procedure faster (especially important with~GPU).

\begin{algorithm}[htb!]
\caption{Recursive procedure $\mathit{MarkSupersets}$.}
\label{alg:marksuper}
\begin{algorithmic}[1]
\Require Lexicographically sorted list intervals $A$ and $B$ of sets with unique elements. Current depth of recursion $d$ ($d=0$ for the initial call).
\Procedure{MarkSupersets}{$A$, $B$, $d$}
\If{$|A| < \mathit{MIN}$} \Comment{Brute-force for small $|A|$}
	\State Check each pair in $A \times B$ and delete the supersets from $B$ (move at the end and shrink the interval).
\Else
\State $A_0, A_1 \gets$ $A$ split by the $d$-th bit \Comment{$A$ is sorted, so a binary search suffices}
\State \Call{$\mathit{MarkSupersets}$}{$A_0, B, d+1$}
\State Sort $B$ by the $d$-th bit. \Comment{Linear time scan}
\State $B_1 \gets$ interval in which the $d$-th bit is set \Comment{Suffix of $B$}
\State \Call{$\mathit{MarkSupersets}$}{$A_1, B_1, d+1$}
\EndIf
\EndProcedure
\end{algorithmic}
\end{algorithm}

$\mathit{MarkSupersets}$ is used to reduce the BFS list and to check the meet condition.
To implement the $\mathit{MarkSubsets}$ procedure needed on the IBFS side, we simply convert the sets to their complements and call $\mathit{MarkSupersets}$.
The procedure $\mathit{MarkProperSupersets}$ is identical except for checking the containment for a pair of sets, where we additionally check that the two sets are different.
When we use multithreading, we split the $B$ list into equal parts after shuffling and execute parallel calls of the procedure.
On GPU, we increase the \textit{MIN} parameter and run the brute-force part there.

\subsubsection{Decision Procedure}

As the algorithm progresses, some steps may become unprofitable.
The history lists, though helpful at the beginning, increase memory usage and cause a slow down if used in late iterations.
Similarly, list reductions via $\mathit{MarkSupersets}$ decrease the branching factor, but they are not that crucial when the search is approaching the reset threshold upper bound.

We distinguish five types of steps from which the algorithm always chooses one for the next iteration -- DFS, BFS, IBFS, BFS without the history list (denoted by $\mathrm{BFS}^\mathrm{NH}$) and IBFS without the history list (denoted by $\mathrm{IBFS}^\mathrm{NH}$).
To assess which option to choose, we roughly estimate the cost subset checking operations each of them will require.

We reuse some of the equations previously defined in~\cite{KKSS2015ComputingTheShortestResetWords}.
In particular, under simplifying assumptions about the uniform distribution of the states in sets we take their upper bound from~\cite[Theorem~4]{KKSS2015ComputingTheShortestResetWords} previously applied to tries.
Since our procedure can be interpreted as building a trie implicitly on the fly, this bound can also serve as a rough bound on the expected number of subset checking operations in a call to $\mathit{MarkSupersets}$.
Let $A_s$, $B_s$ be the size of the lists and $A_d$, $B_d$ be their densities, i.e., for a list $L$ let $\mathit{density}(L) = \frac{\sum_{S \in L}{|S|}}{n |L|}$.
Then, analogously to \cite[\textsc{ExpNvn}~in~4.2]{KKSS2015ComputingTheShortestResetWords}, we define:
\begin{align*}
&& \mathit{ExpMark}&(A_s,B_s, A_d, B_d) = B_s \Big(\frac{1 + B_d}{B_d} + \frac{1}{A_d - A_d B_d}\Big)A_s^{\log_w(1 + B_d)} ,
\end{align*}
where $w = (1+B_d)/(1 + A_d B_d - A_d)$.

To estimate the sizes of the lists after (and in between) the reductions, we store the ratios $r_\mathrm{BFS}^\mathit{dupl}, r_\mathrm{BFS}^\mathit{self}, r_\mathrm{BFS}^\mathit{hist}$ of the reduced sets respectively by the removal of duplicates, the removal of non-minimal subsets, and the reduction by history list.
For instance, $r_\mathrm{BFS}^\mathit{dupl}$ is the fraction of the removed duplicates during the first reduction.
Similarly, separate ones are stored for the IBFS counterpart.
In addition to the cost of subset checking, we also add the cost of computing sets themselves and reduction of duplicates (the constant $\mathit{SetCost}$, set to $512$ in the implementation); however, in most cases, the cost of subset checking is dominant.

The cost of the next BFS step is calculated as follows:
\begin{align}
\mathit{BFS_{cost}} =&\ \mathit{SetCost} \cdot k \cdot |L_\mathrm{BFS}| \label{eq1}\\
&+ \mathit{ExpMark}(k\cdot(1 - r_\mathrm{BFS}^\mathit{dupl})\cdot|L_\mathrm{BFS}|,\quad k\cdot(1 - r_\mathrm{BFS}^\mathit{dupl})\cdot|L_\mathrm{BFS}|, \label{eq2}\\
&\phantom{+ \mathit{ExpMark}(}\ \mathit{density}(L_\mathrm{BFS}),\quad \mathit{density}(L_\mathrm{BFS})) \nonumber\\
&+ \mathit{ExpMark}(|H_\mathrm{BFS}|,\quad k\cdot(1-r_\mathrm{BFS}^\mathit{dupl})\cdot(1-r_\mathrm{BFS}^\mathit{self})\cdot|L_\mathrm{BFS}|, \label{eq3}\\
&\phantom{+ \mathit{ExpMark}(}\ \mathit{density}(H_\mathrm{BFS}),\quad \mathit{density}(L_\mathrm{BFS})) \nonumber\\
&+ \mathit{ExpMark}(k\cdot(1-r_\mathrm{BFS}^\mathit{dupl})\cdot(1-r_\mathrm{BFS}^\mathit{self})\cdot(1-r_\mathrm{BFS}^\mathit{hist})\cdot|L_\mathrm{BFS}|,\quad |L_\mathrm{IBFS}|, \label{eq4}\\
&\phantom{+ \mathit{ExpMark}(}\ \mathit{density}(L_\mathrm{BFS}),\quad \mathit{density}(L_\mathrm{IBFS})). \nonumber
\end{align}
The formula is the sum of the four costs: the set cost (\ref{eq1}), which estimates the cost of computing the successors' list with images and reduction of duplicates, the self-reduction cost (\ref{eq2}), which assumes that the list size was already reduced by the factor $r_\mathrm{BFS}^\mathit{dupl}$, the reduction by history cost (\ref{eq3}), which assumes both preceding reductions, and the meet condition check cost (\ref{eq4}).

The formulas for $\mathrm{BFS}^\mathrm{NH}$, $\mathrm{IBFS}$, and $\mathrm{IBFS}^\mathrm{NH}$ are analogous and as follows:
\begin{align*}
\mathit{BFS_{cost}} =&\ \mathit{SetCost} \cdot k \cdot |L_\mathrm{BFS}| \\
&+ \mathit{ExpMark}(k\cdot(1 - r_\mathrm{BFS}^\mathit{dupl})\cdot|L_\mathrm{BFS}|,\quad k\cdot(1 - r_\mathrm{BFS}^\mathit{dupl})\cdot|L_\mathrm{BFS}|, \\
&\phantom{+ \mathit{ExpMark}(}\ \mathit{density}(L_\mathrm{BFS}),\quad \mathit{density}(L_\mathrm{BFS})) \\
&+ \mathit{ExpMark}(|H_\mathrm{BFS}|,\quad k\cdot(1-r_\mathrm{BFS}^\mathit{dupl})\cdot(1-r_\mathrm{BFS}^\mathit{self})\cdot|L_\mathrm{BFS}|, \\
&\phantom{+ \mathit{ExpMark}(}\ \mathit{density}(H_\mathrm{BFS}),\quad \mathit{density}(L_\mathrm{BFS})) \\
&+ \mathit{ExpMark}(k\cdot(1-r_\mathrm{BFS}^\mathit{dupl})\cdot(1-r_\mathrm{BFS}^\mathit{self})\cdot(1-r_\mathrm{BFS}^\mathit{hist})\cdot|L_\mathrm{BFS}|,\quad |L_\mathrm{IBFS}|, \\
&\phantom{+ \mathit{ExpMark}(}\ \mathit{density}(L_\mathrm{BFS}),\quad \mathit{density}(L_\mathrm{IBFS})).
\end{align*}
\begin{align*}
\mathit{BFS_{cost}^{NH}} =&\ \mathit{SetCost} \cdot k \cdot |L_\mathrm{BFS}| \\
&+ \mathit{ExpMark}(k\cdot(1-r_\mathrm{BFS}^\mathit{dupl})\cdot|L_\mathrm{BFS}|,\quad k\cdot(1 - r_\mathrm{BFS}^\mathit{dupl})\cdot|L_\mathrm{BFS}|, \\
&\phantom{+ \mathit{ExpMark}(}\ \mathit{density}(L_\mathrm{BFS}),\quad \mathit{density}(L_\mathrm{BFS})) \\
&+ \mathit{ExpMark}(k\cdot(1-r_\mathrm{BFS}^\mathit{dupl})\cdot(1-r_\mathrm{BFS}^\mathit{self})\cdot|L_\mathrm{BFS}|,\quad |L_\mathrm{IBFS}|, \\
&\phantom{+ \mathit{ExpMark}(}\ \mathit{density}(L_\mathrm{BFS}),\quad \mathit{density}(L_\mathrm{IBFS})).
\end{align*}
\begin{align*}
\mathit{IBFS_{cost}} =&\ \mathit{SetCost} \cdot k \cdot |L_\mathrm{IBFS}| \\
&+ \mathit{ExpMark}(k\cdot(1-r_\mathrm{IBFS}^\mathit{dupl})\cdot|L_\mathrm{IBFS}|,\quad k\cdot(1 - r_\mathrm{IBFS}^\mathit{dupl})\cdot|L_\mathrm{IBFS}|, \\
&\phantom{+ \mathit{ExpMark}(}\ 1-\mathit{density}(L_\mathrm{IBFS}),\quad 1-\mathit{density}(L_\mathrm{IBFS})) \\
&+ \mathit{ExpMark}(|H_\mathrm{IBFS}|,\quad k\cdot(1-r_\mathrm{IBFS}^\mathit{dupl})\cdot(1 - r_\mathrm{IBFS}^\mathit{self})\cdot|L_\mathrm{IBFS}|, \\
&\phantom{+ \mathit{ExpMark}(}\ 1-\mathit{density}(H_\mathrm{IBFS}),\quad 1-\mathit{density}(L_\mathrm{IBFS})) \\
&+ \mathit{ExpMark}(|L_\mathrm{BFS}|,\quad k\cdot(1-r_\mathrm{IBFS}^\mathit{dupl})\cdot(1 - r_\mathrm{IBFS}^\mathit{self})\cdot(1 - r_\mathrm{IBFS}^\mathit{hist})\cdot|L_\mathrm{IBFS}|, \\
&\phantom{+ \mathit{ExpMark}(}\ \mathit{density}(L_\mathrm{BFS}), \mathit{density}(L_\mathrm{IBFS})).
\end{align*}
\begin{align*}
\mathit{IBFS_{cost}^{NH}} =&\ \mathit{SetCost} \cdot k \cdot |L_\mathrm{IBFS}| \\
&+ \mathit{ExpMark}(k\cdot(1-r_\mathrm{IBFS}^\mathit{dupl})\cdot|L_\mathrm{IBFS}|,\quad k\cdot(1-r_\mathrm{IBFS}^\mathit{dupl})\cdot|L_\mathrm{IBFS}|, \\
&\phantom{+ \mathit{ExpMark}(}\ 1-\mathit{density}(L_\mathrm{IBFS}),\quad 1-\mathit{density}(L_\mathrm{IBFS})) \\
&+ \mathit{ExpMark}(|L_\mathrm{BFS}|,\quad k\cdot(1-r_\mathrm{IBFS}^\mathit{dupl})\cdot(1-r_\mathrm{IBFS}^\mathit{self})\cdot|L_\mathrm{IBFS}|, \\
&\phantom{+ \mathit{ExpMark}(}\ \mathit{density}(L_\mathrm{BFS}),\quad \mathit{density}(L_\mathrm{IBFS})).
\end{align*}

Additionally, if we cannot perform a step because there is not enough memory, we set its expected cost to $\infty$.
Once we choose $\mathrm{BFS}^\mathrm{NH}$, we free the history and no longer consider the BFS step with it, so in this case, we also set its cost to $\infty$ (this is symmetrical for $\mathrm{IBFS}^\mathrm{NH}$ and $\mathrm{IBFS}$).

Next, we try to predict the full costs of choosing these steps by estimating the number of operations under the assumption that the algorithm will transition into the DFS phase one iteration later (or in the current iteration in the case of the $\mathit{DFS}$ option).
First, we calculate the expected branching factor in the DFS phase
\[ f = k \cdot (1 - r_\mathrm{IBFS}^\mathit{dupl} \cdot \mathit{DFSReductionOfReduction}), \]
where we take the average reduction of duplicates from the IBFS step and reduce it by some factor for a more pessimistic estimation ($\mathit{DFSReductionOfReduction} = 1/k$ in the implementation).
$r_\mathrm{IBFS}^\mathit{dupl}$ is the ratio of duplicates removed in the IBFS list during the latest reduction (since the DFS also removes duplicates).
We assume conservatively that the reset threshold is equal to the known upper bound $R$ and from that, we get the estimated number of iterations $R-r$ that still need to be done, where $r$ is the number of the current iteration, which just begins.
The predicted full cost of the BFS option is as follows (the number of sets multiplied by the set computing costs together with the meet condition cost):
\begin{align*}
\mathit{DFS_{pred}} =&\ f \cdot \frac{f^{R - r} - 1}{f - 1} \\
&\cdot(\mathit{SetCost} \cdot \mathit{DFSSetCostWeight} \cdot K / f \\
&\phantom{\cdot(}\ + \mathit{DFSCheckCostWeight} \\
&\phantom{\cdot(+}\ \cdot \mathit{ExpMark}(|L_\mathrm{BFS}|,\quad |L_\mathrm{IBFS}|,\quad \mathit{density}(L_\mathrm{BFS}),\quad \mathit{density}(L_\mathrm{IBFS}))\ ).
k\end{align*}
where $\mathit{DFSSetCostWeight}$ and $\mathit{DFSCheckCostWeight}$ are constants (both set to $0.25$ in the implementation) compensating for the fact that the operations are faster in the DFS phase, as additional optimizations are possible.

The predicted costs of the four other options are analogous and as follows:
\begin{align*}
\mathit{BFS_{pred}} =&\ \mathit{BFS_{cost}} + f \cdot \frac{f^{R - r - 1} - 1}{f - 1} \\
&\cdot(\mathit{SetCost} \cdot \mathit{DFSSetCostWeight} \cdot  / f \\
&\phantom{\cdot(}\ + \mathit{DFSCheckCostWeight} \\
&\phantom{\cdot(+}\ \cdot \mathit{ExpMark}(k\cdot(1-r_\mathrm{BFS}^\mathit{dupl})\cdot(1-r_\mathrm{BFS}^\mathit{self})\cdot(1-r_\mathrm{BFS}^\mathit{hist}) \cdot |L_\mathrm{BFS}|, \, |L_\mathrm{IBFS}|, \\
&\phantom{\cdot(+\ \cdot \mathit{ExpMark}(}\mathit{density}(L_\mathrm{BFS}),\quad \mathit{density}(L_\mathrm{IBFS})).
\end{align*}
\begin{align*}
\mathit{BFS_{pred}^{NH}} =&\ \mathit{BFS_{cost}^{NH}} + f \cdot \frac{f^{R - r - 1} - 1}{f - 1} \\
&\cdot(\mathit{SetCost} \cdot \mathit{DFSSetCostWeight} \cdot K / f \\
&\phantom{\cdot(}\ + \mathit{DFSCheckCostWeight} \\
&\phantom{\cdot(+}\ \cdot \mathit{ExpMark}(k\cdot(1-r_\mathrm{BFS}^\mathit{dupl})\cdot(1-r_\mathrm{BFS}^\mathit{self}) \cdot |L_\mathrm{BFS}|, \, |L_\mathrm{IBFS}|, \\
&\phantom{\cdot(+\ \cdot \mathit{ExpMark}(}\mathit{density}(L_\mathrm{BFS}),\quad \mathit{density}(L_\mathrm{IBFS})).
\end{align*}
\begin{align*}
\mathit{IBFS_{pred}} =&\ \mathit{IBFS_{cost}} + f \cdot \frac{f^{R - r - 1} - 1}{f - 1} \\
&\cdot(\mathit{SetCost} \cdot \mathit{DFSSetCostWeight} \cdot K / f \\
&\phantom{\cdot(}\ + \mathit{DFSCheckCostWeight} \\
&\phantom{\cdot(+}\ \cdot \mathit{ExpMark}(|L_\mathrm{BFS}|, \, k\cdot(1-r_\mathrm{IBFS}^\mathit{dupl})\cdot(1-r_\mathrm{IBFS}^\mathit{self})\cdot(1-r_\mathrm{IBFS}^\mathit{hist}) \cdot |L_\mathrm{IBFS}|,\\
&\phantom{\cdot(+\ \cdot \mathit{ExpMark}(}\mathit{density}(L_\mathrm{BFS}),\quad \mathit{density}(L_\mathrm{IBFS})).
\end{align*}
\begin{align*}
\mathit{IBFS_{pred}^{NH}} =&\ \mathit{IBFS_{cost}^{NH}} + f \cdot \frac{f^{R - r - 1} - 1}{f - 1} \\
&\cdot(\mathit{SetCost} \cdot \mathit{DFSSetCostWeight} \cdot K / f \\
&\phantom{\cdot(}\ + \mathit{DFSCheckCostWeight} \\
&\phantom{\cdot(+}\ \cdot \mathit{ExpMark}(|L_\mathrm{BFS}|, \, k\cdot(1-r_\mathrm{IBFS}^\mathit{dupl})\cdot(1-r_\mathrm{IBFS}^\mathit{self}) \cdot |L_\mathrm{IBFS}|,\\
&\phantom{\cdot(+\ \cdot \mathit{ExpMark}(}\mathit{density}(L_\mathrm{BFS}),\quad \mathit{density}(L_\mathrm{IBFS})).
\end{align*}

Finally, the step with the lowest predicted full cost is chosen.
As an exception to this rule, if we do not expect to switch into the DFS phase soon (the steps without the history have both larger costs than the regular steps), instead of comparing the prediction costs of every choice, we consider only the BFS and IBFS costs and greedily choose the one with the lower single-step cost.
This makes the bidirectional search more balanced in the short term.
Otherwise, BFS is strongly preferred due to the assumption that the rest of the steps will be (inverse) DFS, which computes preimages, but the early statistics are less relevant for its estimation.

\subsubsection{Heuristic Upper Bound}

Having a good upper bound $R$ on the reset threshold, preferably tight, is crucial for making the right decisions, i.e., not giving away history or entering the DFS phase too late.
On the other hand, we do not want to spend too much time computing the bound.

The beam algorithm has its parameter \emph{beam size}, which limits the size of the list and thus directly controls the quality (i.e., how close $R$ is to the reset threshold) and complexity trade-off.
Instead of setting the beam size to be a function of $n$ (\cite{KKSS2015ComputingTheShortestResetWords,RS2015ForwardAndBackward}), we use an adaptive approach.
We run it first with a relatively small beam size to find some reasonable bound, and then use this bound to calculate a rough estimation of the running time of the exact algorithm.
The beam size is selected so that the beam algorithm's cost is a small fraction of that of the exact algorithm.
In practice, this means that the beam size is larger for automata with larger (upper bounds on) reset thresholds than for those automata with smaller ones, even when $n$ is the same.

\subsection{Depth-First Search}

In the second phase, the algorithm switches to an inverse depth-first search, which allows staying within the memory limit by adjusting the maximum list size.
During this phase, the steps are taken only on the IBFS side.
The fact that the BFS list no longer changes allows the meet condition check to be optimized.

\subsubsection{Static Radix Trie}

The $L_\mathrm{BFS}$ list is stored in an optimized data structure that supports the $\mathit{ContainsSubset}$ operation.
It is based on a radix trie, in which the characteristic vectors of the sets are stored.
The queries rely on traversing the trie and, in each step, descending to either both children or just the left (zero) child, depending on the queried set.

In contrast with the usual radix trie, we apply a few specific optimizations:

\noindent$\bullet$\ \textbf{Variable state ordering}.
In each node at a depth $d$, instead of splitting the subtrees by the $d$-th state, we split by a state $x$ chosen specifically for the node.
The state $x$ is chosen so that the number of sets stored in the current subtree that also contain $x$ is the largest possible.
In practice, this makes the queries faster, since, in vertices such that $x$ does not belong to the queried set, a large number of sets is immediately skipped.

This optimization also implies path compression, as we do not have nodes that do not split the sets into two non-empty parts.
To ensure this, we also exclude the case that $x$ is contained in all the sets and do not use it as a division state.

\noindent$\bullet$\ \textbf{Leaf threshold}.	
For a fixed constant parameter $\mathit{MIN}$ ($=10$ in the implementation), when the number of sets in a subtree is less or equal to $\mathit{MIN}$, we store them all in one vertex and do not recurse further.
This does not increase running time and lowers the memory overhead.
Technically, we can already store the sets in a node whose left (zero) subtree contains at most $\mathit{MIN}$ sets, which avoids creating an additional node.

\noindent$\bullet$\ \textbf{Joint queries}.
Instead of checking each subset separately, we group them by the same cardinality and check one group in a call using the swapping technique as in Alg.~\ref{alg:marksuper}.
Grouping saves operations responsible for traversing the trie, and additionally, grouping by cardinality allows to make cheap size elimination checks as below.

\noindent$\bullet$\ \textbf{Size elimination}.
Every node $v$ stores the minimal size $m_v$ of the sets in its subtree.
When we query for subsets of size $s$, if $m_v \leq s$ does not hold, we do not recurse into the subtree of $v$.
This enhancement is derived from the previous algorithm \cite{KKSS2015ComputingTheShortestResetWords}, yet due to joint queries, we make these checks cheaper by executing at most one such check in a node for one cardinality.
A similar optimization from the previous algorithm is mask elimination -- checking if the intersection of all subsets in the subtree is contained in the queries set, yet we do not use it as it does not improve performance in our case.

Our trie can be built in time $\O(|L_\mathrm{BFS}| \cdot n^2)$ ($< n$ levels processed in time $\O(|L_\mathrm{BFS}| \cdot n$)).

\subsection{DFS Procedure}

During the search, at each depth, the current list is split into parts of size at most $\mathit{available\_memory} / \big((k + 1)(R-r)\big)$, where $R-r$ is the upper bound on the remaining number of steps to be done.
Then, it recurses with each of these parts one by one, to make sure we do not run out of memory.
The~elements are sorted in order of descending cardinality so that the most promising sets are recursed first, which in turn can quickly improve the upper bound if it was not tight.
The cardinality sort is also necessary for joint queries.
The lists are reduced by the removal of duplicates and calls to $\mathit{MarkSubsets}$ only once every few iterations, which still lowers the branching factor significantly.
Parallel computation is performed by a thread pool with tasks being these separate calls for each cardinality.

The recursive DFS procedure is shown in~Alg.~\ref{alg:dfs}.

\begin{algorithm}[htb]
\caption{Recursive procedure $\mathit{DFS}$.}\label{alg:dfs}
\begin{algorithmic}[1]
\Require Static radix trie $\mathit{T}$ built over the elements of $L_\mathrm{BFS}$.
List $L$ -- initially $L_\mathrm{IBFS}$.
The current reset word length $r$ (BFS iterations + recursion depth).
The reset threshold upper bound $R$, which can be changed globally once we find a shorter reset word. A fixed parameter $C$, set to $20{,}000$ in our implementation.
\Ensure $R$ will be set to match the reset threshold.
\State $L_\mathit{next} \gets$ \Call{$\mathit{CalculatePreimages}$}{$L$}
\State Sort $L_\mathit{next}$ in the order of decreasing cardinality
\If{Time for duplicates removal}
\State Remove duplicates from $L_\mathit{next}$
\EndIf
\If{Time for reduction}
\State Delete subsets of the $C$ largest sets in $L_\mathit{next}$ ($\textit{MarkSubsets}$)
\EndIf
\For{$c \in \{2,\ldots,N-1\}$} \Comment{For each cardinality}
\State $L_c \gets$ sublist of $L_\mathit{next}$ of sets with cardinality $c$
\If{$T$ contains subset of a set in $L_c$}
\State $R \gets r$
\State \Return
\EndIf
\EndFor
\If{$r = R - 1$}
	\State \Return \Comment{Failed to find a reset word shorter than $R$}
\EndIf
\For{$L_\mathit{part} \in \mathit{Split}(L_\mathit{next}$)} \Comment{Split into smaller parts if necessary}
\State \Call{$\mathit{DFS}$}{$T, L_\mathit{part}, r + 1, R$}
\If{$r = R - 1$}
	\State \Return
\EndIf
\EndFor
\end{algorithmic}
\end{algorithm}

\section{Experiments}

The implementation used for experiments is available at~\cite{SZ2022Synchrowords}.
The tests were using \texttt{exact\_reduce} configuration.
The \emph{old} algorithm \cite{KKSS2015ComputingTheShortestResetWords} was run with the original code provided to us by the authors.
The experiments with time measurement were run on computers with AMD Ryzen Threadripper 3960X 24-Core Processor, 64GB RAM, and two RTX3080 Nvidia GPU cards.
We compiled the code using \texttt{gcc 9.3.0} and \texttt{nvcc 10.1} (run with \texttt{gcc 7.5.0}).
We have tested the algorithm in both the \emph{single-threaded} (only one thread without GPU) and \emph{parallel} modes (6 threads and GPU enabled).

A \emph{random} automaton with $n$ states and $k$ letters is generated by choosing each transition $\delta(q,a) \in Q$ uniformly at random, for $q \in Q$, $a \in \Sigma$.
There is a negligible number of non-synchronizing automata obtained in this way, which were excluded (cf.\ \cite{Berlinkov2016OnTheProbabilityToBeSynchronizable}).

\subsection{The Efficiency}

\begin{figure}[H]\centering
\includegraphics[width=.9\textwidth]{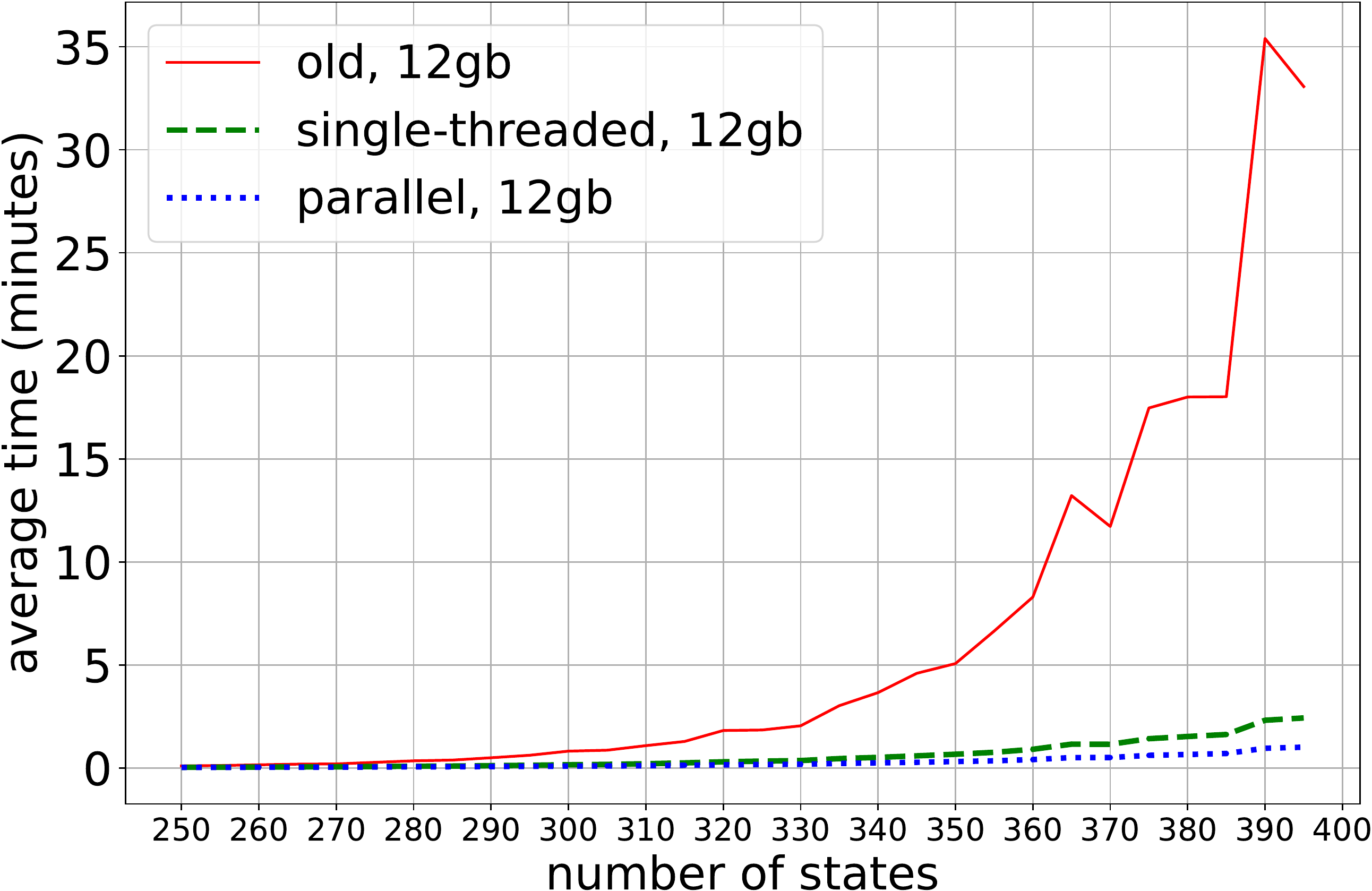}
\caption{The mean running time for random binary automata with different numbers of states.}
\label{fig:efficiency_n}
\end{figure}

\begin{figure}[H]\centering
\includegraphics[width=.9\textwidth]{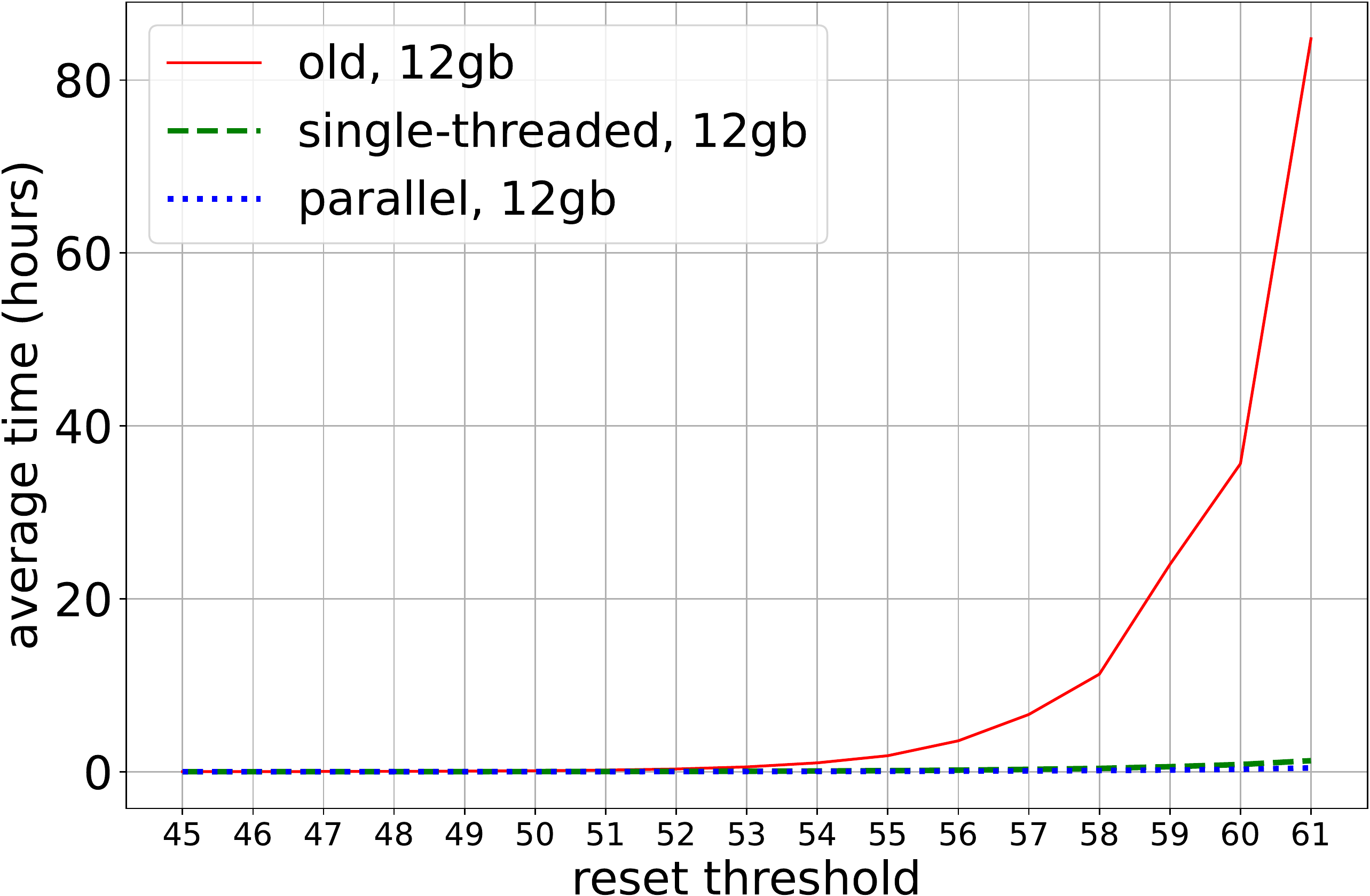}
\caption{The mean running time for random binary automata with different reset thresholds.}
\label{fig:efficiency_rt}
\end{figure}

\begin{figure}[H]\centering
\includegraphics[width=.9\textwidth]{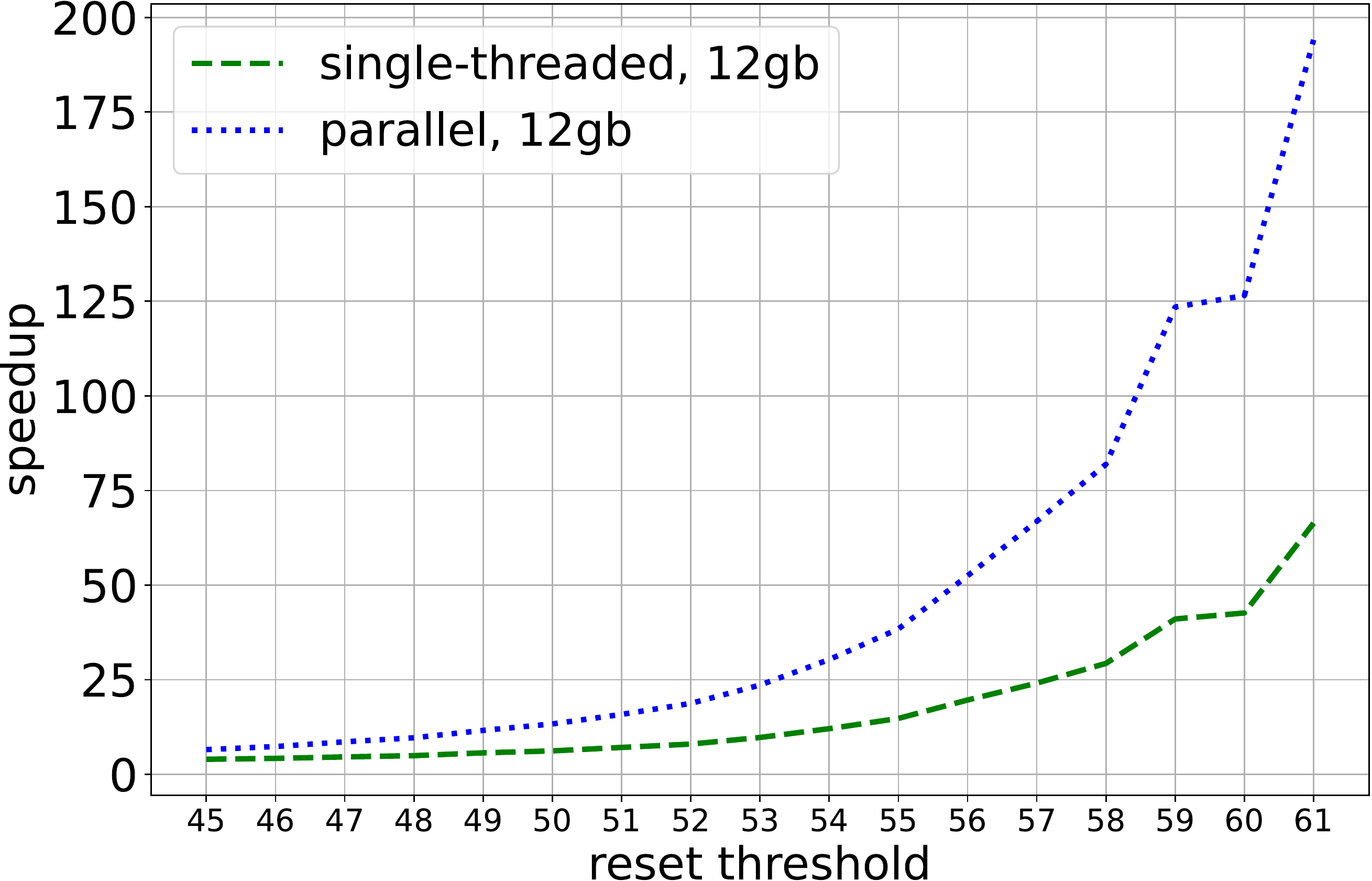}
\caption{The average speed-up in relation to the old algorithm (the running time of our algorithm divided by the running time of the old one).}
\label{fig:speedup}
\end{figure}

Fig.~\ref{fig:efficiency_n} and~\ref{fig:efficiency_rt} show the efficiency comparison of our algorithm in both single-threaded and parallel modes, together with the old algorithm.
This experiment was run for $1{,}000$ random binary automata for each $n \in \{250,255,\ldots,395\}$.
We managed to test automata with up to $395$ with the old algorithm, which took $3{,}177$h  computation time of a single process in total (we have used more memory, better hardware, and computed 10 times fewer automata per $n$ than in the original experiments \cite{KKSS2015ComputingTheShortestResetWords}, which were done up to $n=350$).
The (unfinished) attempt to compute $1,000$ automata with $n=400$ by the old algorithm took over $770$h, whereas our algorithm (single-threaded) finished in $48$h.
The total computation time of our algorithm up to $n=395$ was resp.\ $299$h in the single-threaded mode and $135$h in the parallel mode.
Fig.~\ref{fig:speedup} shows the mean running time of our algorithm in relation to the running time of the old one.

\begin{figure}[H]\centering
\includegraphics[width=.9\textwidth]{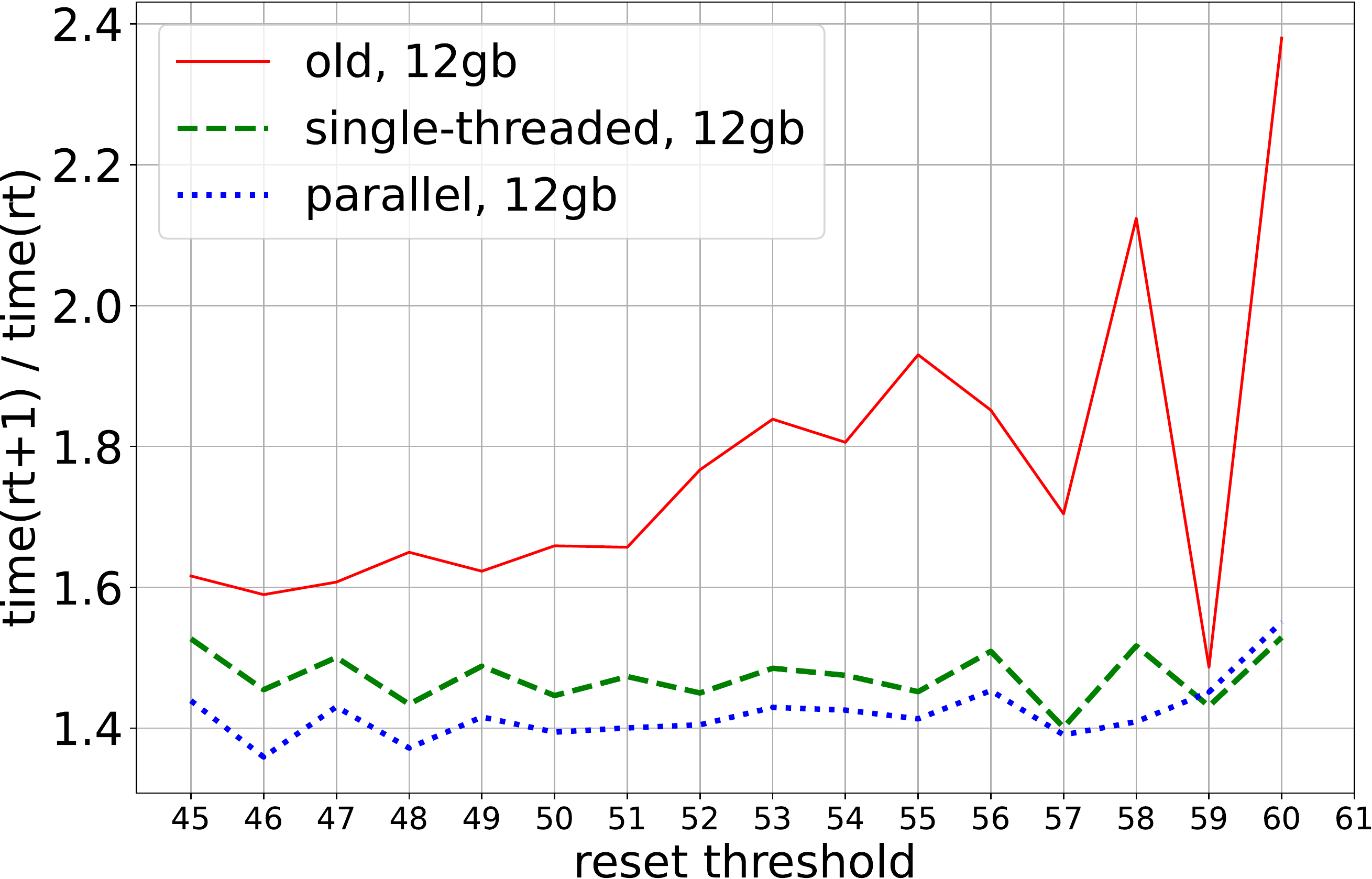}
\caption{The mean running time growth factor in relation to reset threshold.}
\label{fig:efficiency_growth}
\end{figure}

\begin{figure}[H]\centering
\includegraphics[width=.9\textwidth]{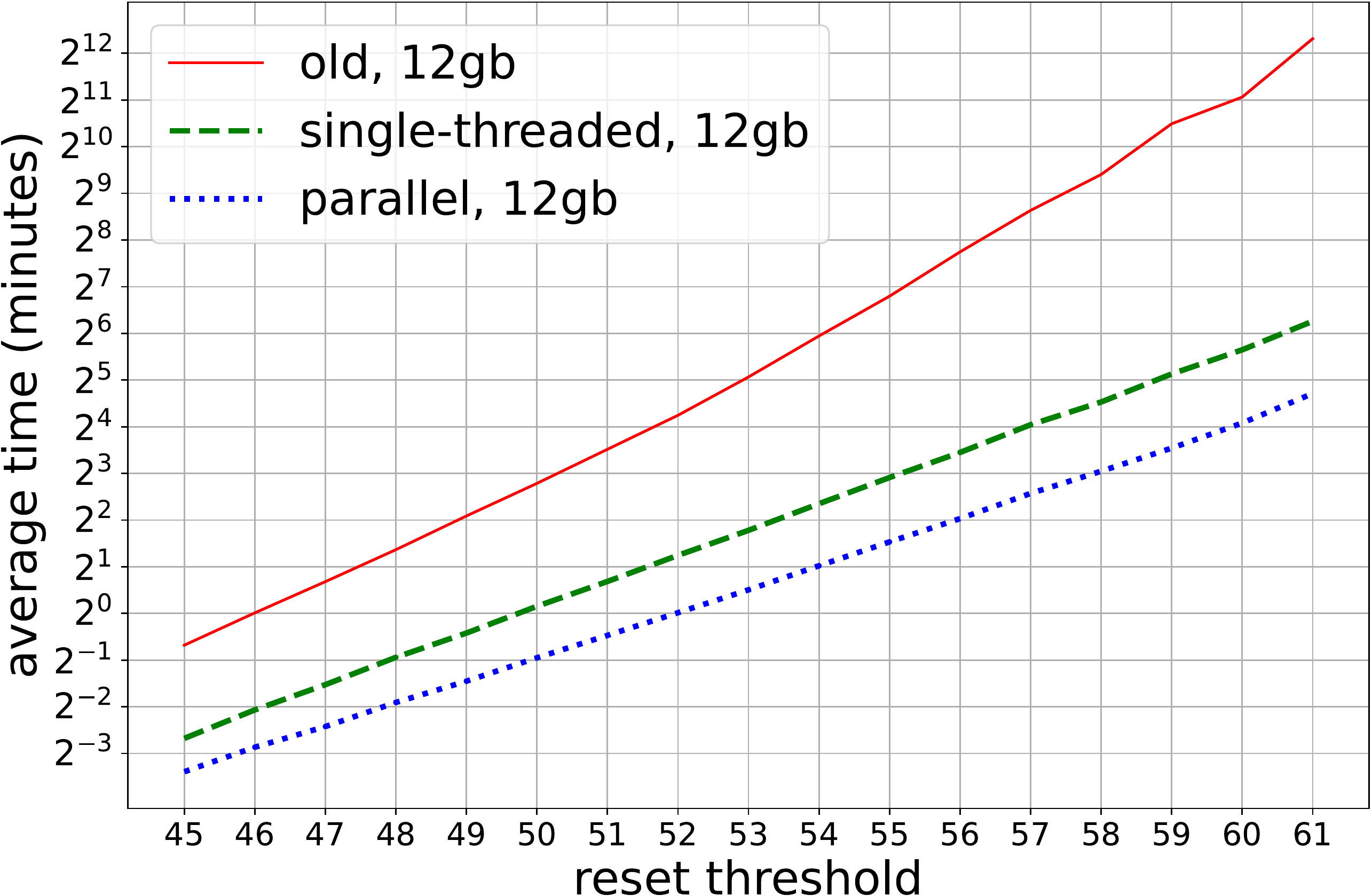}
\caption{The mean running time growth factor in relation to reset threshold in a logarithmic scale.}
\label{fig:efficiency_growth_log}
\end{figure}

Fig.~\ref{fig:efficiency_growth} and~\ref{fig:efficiency_growth_log} show the average increase in running time when the reset threshold grows.
The general observed tendency is a factor of about $1.5$ for our algorithm.

\subsubsection{Hard instances}

Most automata have their reset thresholds sublinear, but there exist other examples (though they are rare).
As the reset threshold is the main indicator of difficulty, even instances with a small number of states should be difficult for algorithms.

From the known constructions, the most extreme automata with respect to the reset threshold are \emph{slowly synchronizing} ones
\cite{AGV2013SlowlySynchronizing}.
They have reset thresholds close to $(n-1)^2$, and the \Cerny series meets this bound.
Yet, they all have the property that the IBFS list reduces to a constant number of sets in every iteration, thus our algorithm works in polynomial time for them, just as the old algorithm.
An example of extreme automata without this property could be the unique series of automata with a \emph{sink} state (i.e., a state $q \in Q$ such that $\delta(q,a)=q$ for all $a \in \Sigma$), that reaches the maximum reset threshold $n(n-1)/2$ in this class \cite{R1997ResetWordsForCummutativeAndSolvableAutomata}.
(A synchronizing automaton can have at most one sink state and a reset word must map all the states to it.)
A \emph{slowly sink} automaton from this series with $26$~states has $25$~letters and its reset threshold $325$ is computed by the old algorithm in $34$m $14$s, whereas the new algorithm computes it in $7$m $22$s (parallel).

\subsection{Mean reset threshold}

\begin{figure}[H]\centering
\includegraphics[width=.9\textwidth]{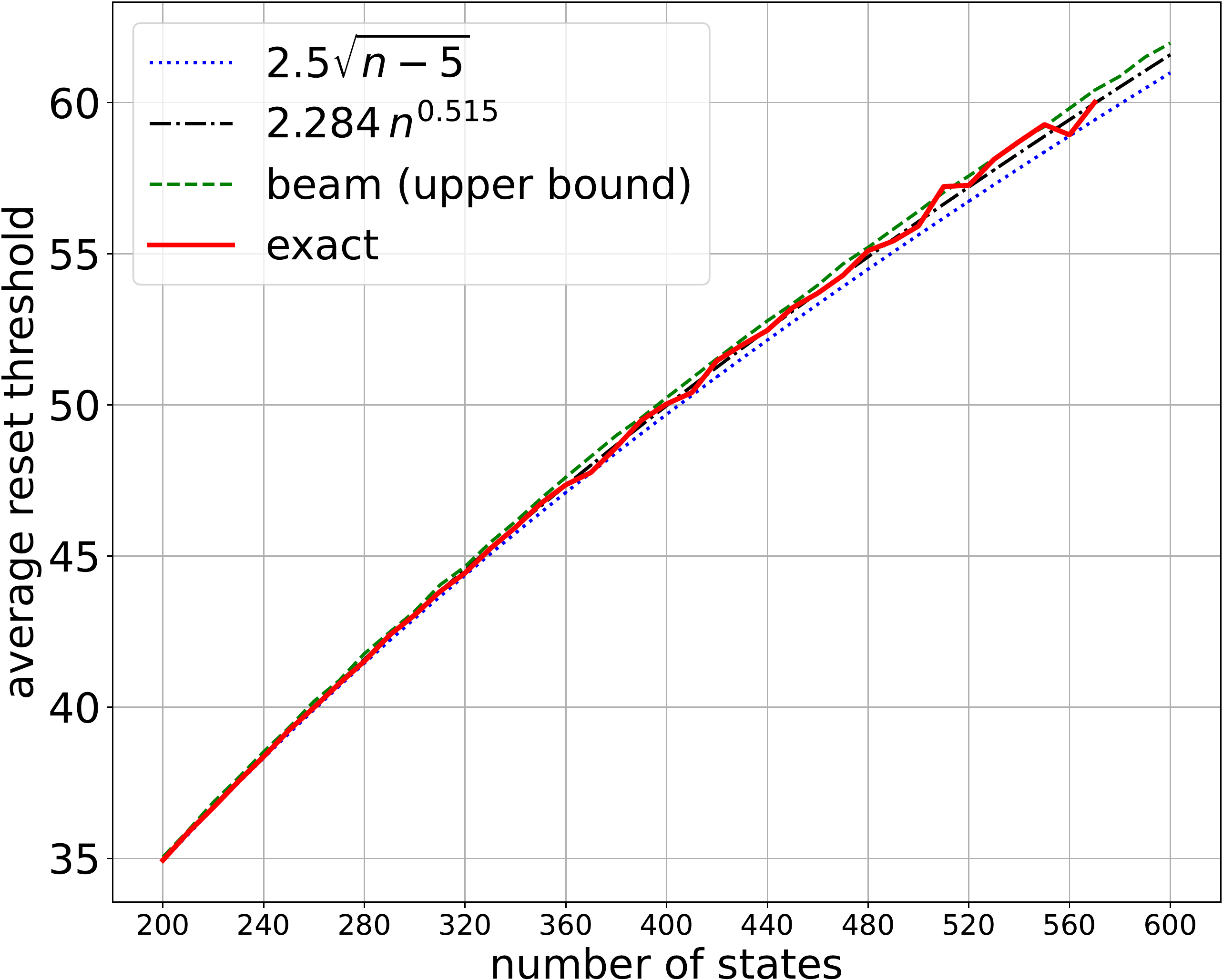}
\caption{The mean reset threshold of binary automata with $n$ states.
For every $n \in \{5,10,\ldots,300\}$, $n \in \{305,310,\ldots,400,410,420,\ldots,500\}$, and $n \in \{510, 520,\ldots,570\}$ we calculated resp.\ $10{,}000$, $1{,}000$, and $100$ automata.}
\label{fig:mean_rt}
\end{figure}

\begin{figure}[H]\centering
\includegraphics[width=.9\textwidth]{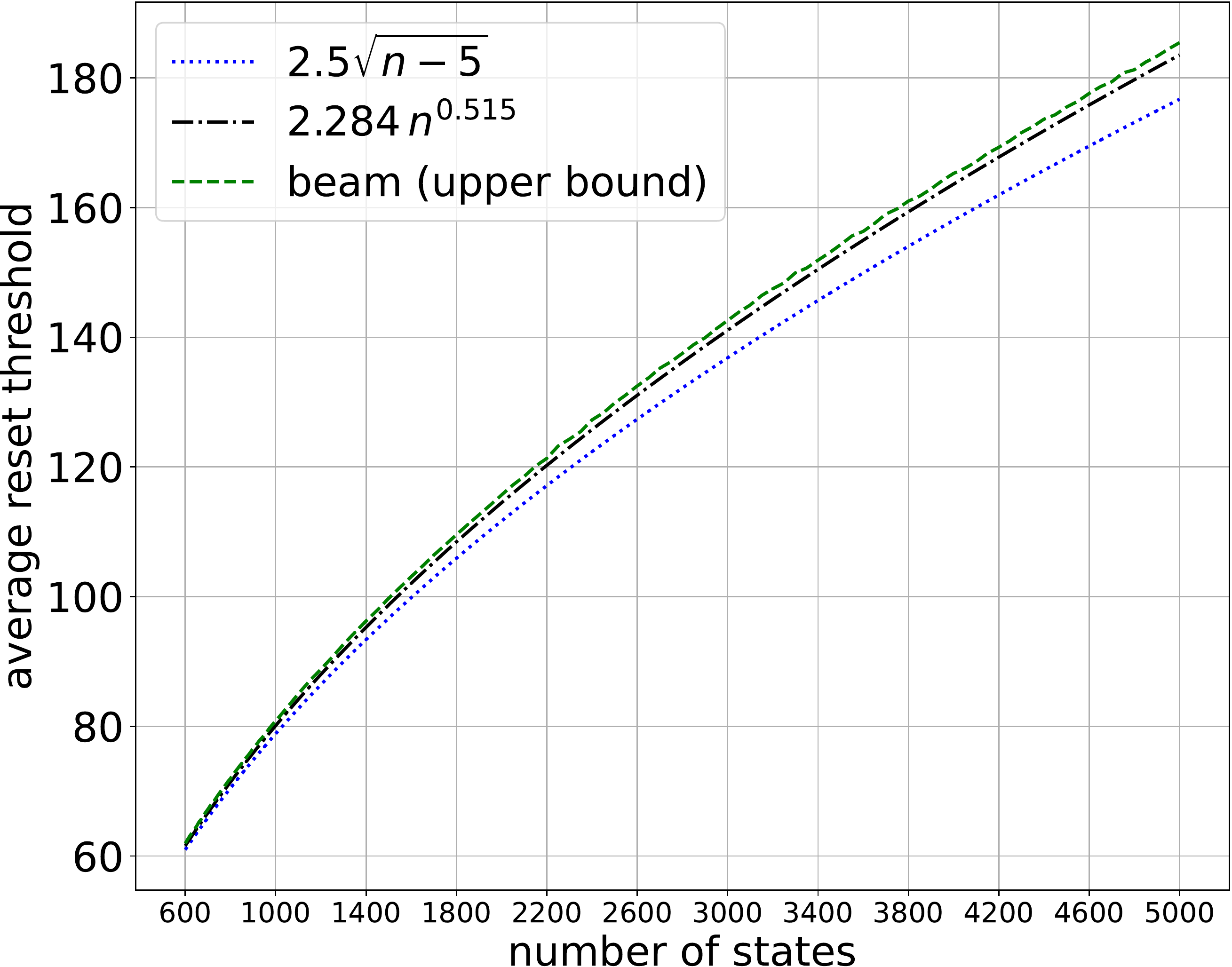}
\caption{The upper bound on the reset threshold of binary automata with $n$ states.
For every $n \in \{100,105,\ldots, 1000, 1050, \ldots, 2000\}$, $n \in \{2050,2100,\ldots,5000\}$ we calculated resp.\ $10{,}000$ and $1{,}000$ automata with the beam algorithm with beam size $n \log n$.}
\label{fig:mean_rt_beam}
\end{figure}

In the second experiment, we computed reset thresholds of random binary automata with $n \in \{410,420,\ldots,570\}$ states with our algorithm in parallel mode.
The mean computation time for $n=500$ was $9$m $42$s.
Fig.~\ref{fig:mean_rt} shows the mean reset threshold.
In addition, up to $n=5{,}000$ we computed a good upper bound using the beam algorithm with beam size $n \log n$ (we made the beam algorithm much faster due to GPU computation; the previous such experiments were done up to $n=1000$ \cite{RS2015ForwardAndBackward}); see Fig.~\ref{fig:mean_rt_beam}.

\begin{figure}[htb]\centering
\includegraphics[width=.9\textwidth]{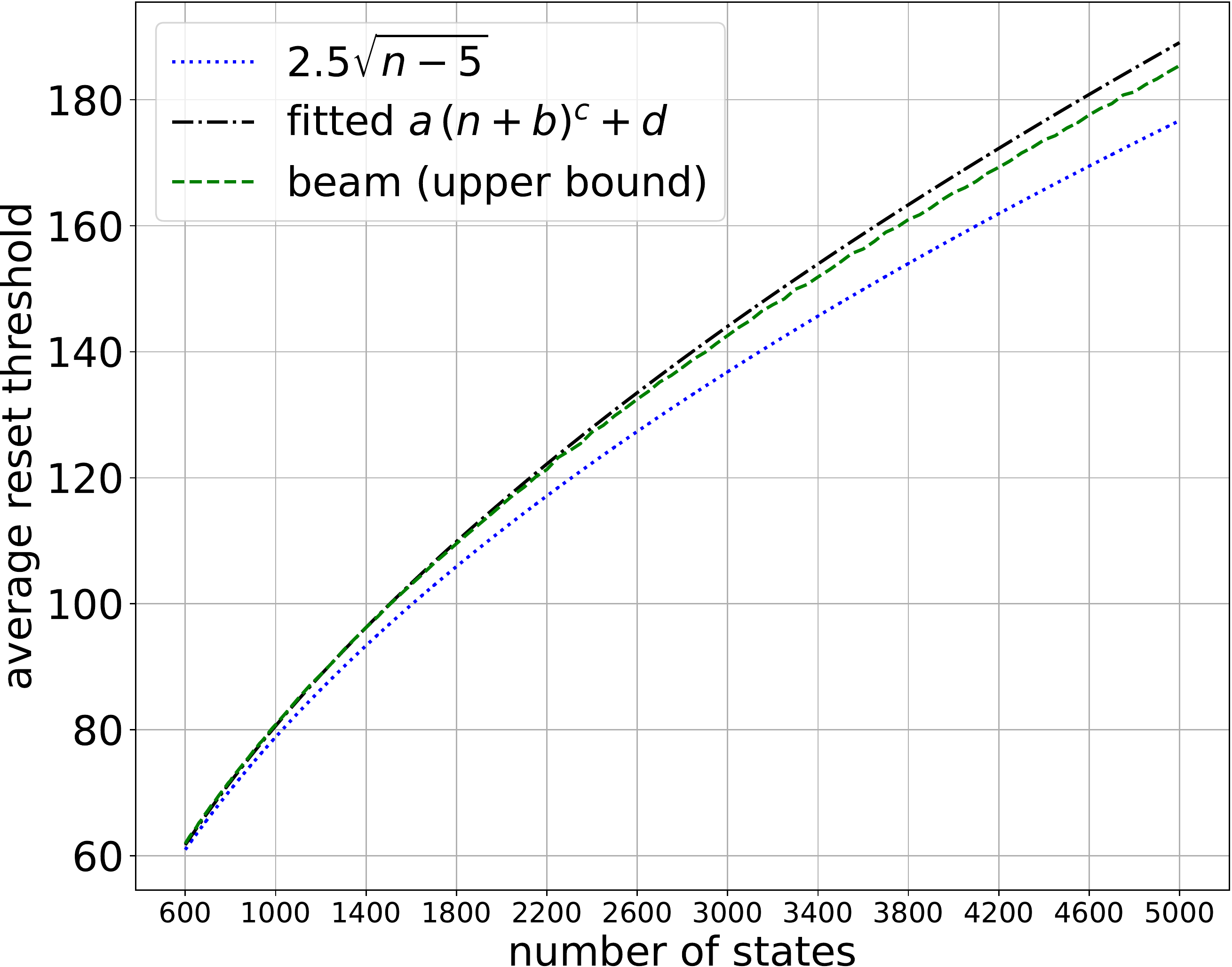}
\caption{The mean reset threshold with a wrong estimation obtained directly from the results of the exact algorithm.}
\label{fig:wrong_fit}
\end{figure}

It is now visible that the previous formula $2.5\sqrt{n-5}$ is underestimated.
On the other hand, a standard approach\footnote{We use the algorithm from \texttt{scipy.optimize.curve\_fit} with the Levenberg-Marquardt algorithm.} of deriving a function of the form $a(n+b)^c+d$ yields a wrong formula, lately exceeding the upper bound obtained by the beam (see Fig.~\ref{fig:wrong_fit}).
We decrease the exponent to fit with a more accurate estimation $2.284\, n^{0.515}$.

\section{Conclusions}

We have improved the best-known algorithm for computing the (length of the) shortest reset words.
While the overall idea of employing bidirectional breadth-first search is the same, we replace each of its subprocedures with more efficient ones.

The algorithm can be easily adapted to alternative synchronization settings and other related problems.
For instance, it is trivial to use it for \emph{careful} synchronization \cite{Shabana2019}, where some transitions can be forbidden.
It can also be adapted to, e.g, non-careful settings \cite{BFRS21SynchronizingStronglyConnectedPartialDFAs}, \emph{mortal words} \cite{R2019MortalityAndSynchronizationOfUnambiguous}, or subset synchronization \cite{V2016SubsetSynchronization}.

For future work, we plan to use this new algorithmic tool to perform more extensive experiments concerning reset thresholds, especially with larger automata and with a larger alphabet.
Finally, it can be used to experimentally verify or extend the current verification range of certain conjectures.

\bibliography{synchronization}
\end{document}